# Bioadhesive Hydrogel Flexible Laser for Sweat Sensing based on Liquid Crystal Microdroplets

*Ningyuan Nie[1], Yu-Cheng Chen[1*]*

[1] School of Electrical and Electronic Engineering, Nanyang Technological University, 50 Nanyang Avenue, 639798, Singapore

Correspondence Email: yucchen@ntu.edu.sg

**Abstract**

Flexible photonics offers the possibility to realize wearable sensors by bridging the advantages of flexible materials and photonic sensing elements. Recently, optical resonators have emerged as a tool to improve its over sensitivity by integrating with flexible photonic sensors. However, direct monitoring of multiple psychological information on human skin remains challenging, due to the subtle biological signals and complex tissue interface. To tackle the current challenges, here we developed a functional thin film laser formed by encapsulating multiple liquid crystal microdroplet laser resonators in a flexible hydrogel for monitoring important metabolites in human sweat (lactate, glucose, and urea). The three-dimensional cross-linked hydrophilic polymer serves as the adhesive layer to allow small molecules to penetrate from human tissue to generate strong light-matter interactions on the interface of whispering gallery modes resonators. Both hydrogel and CLC microdroplets were modified specifically to achieve high sensitivity and selectivity. As a proof-of-concept, wavelength-multiplexed sensing and a prototype were demonstrated on human skin to detect human metabolites from perspiration. These results present a significant advance in the fabrication and potential guidance for wearable and functional microlasers in healthcare.

**Keywords:** Flexible microlasers, microcavity, whispering gallery mode, sweat sensing

## 1. Introduction

Flexible devices have received tremendous attention in the last decade owing to their weight lightweight, bendability, stretchability, and ease of integration with human interfaces[1-6]. By bridging flexible materials with photonics, flexible photonics offered the possibility to realize flexible light sources, flexible displays, solar cells, flexible photodetectors, anticounterfeiting labels, and wearable sensors[7-16]. Recently, wearable photonic sensors have been extensively explored as an alternative to state-of-the-art wearable sensors owing to their resistance to electromagnetic radiation and environmental changes. Optical wearables are also known for their potential capability to perform remote sensing and detection of multiparameters at the same time. To date, optical wearable sensors have been developed to detect humidity, physical motions, respiratory, heart rate, and temperature[7, 8, 17-20]. Various microscale and nanoscale photonic resonators have been incorporated into flexible materials to improve the sensitivity of wearable optical sensors, including plasmonics, Bragg grating, fiber, and whispering gallery mode micro-resonators. For instance, Zhao *et al.*, developed a flexible organic microlaser array which can detect the gesture of human fingers[10]. Choi *et al.*, demonstrated a flexible bandgap nanolaser with a semiconductor slab embedded in polymers, yielding an optical strain sensitivity for mechanical detection[9].Zhai *et al.*, developed various types of flexible lasers from plasmonic fibers and microrings to detect environmental changes on human surface[17, 21].

Direct sensing of biochemicals released in human body can provide more clinical-relevant bioinformation. As a matter of fact, human sweat contains a plethora of biomarkers which provide key physiological information related to human function and metabolism[22-26]. Compared with blood testing, sweat testing offers the advantages of non-invasiveness, portability, and persistence[27-29]. Hence, analysis and detection of biomarkers in sweat can assist in the prevention, diagnosis, and especially monitoring of chronic diseases. Previous studies have investigated the possibility of using surface-enhanced Raman scattering, photonic crystals-based structural color, and polarized microscope for sweat sensing[30-33]. Despite the rapid advancement in wearable optical sensors, one of the greatest challenges is to detect multiple biochemicals on a single device, which means being capable of multiplexed detection or multifunctionality.

To overcome the limitations, stimulated emissions from micro- to nanoscale lasers offer unique advantages in terms of signal amplification and narrow linewidth[34-38]. Strong light interactions between optical microcavities and biomolecules would therefore lead to distinctive lasing signals for sensing[39-44]. In particular, droplet-based microlasers are promising candidates for their compact size, ease of fabrication, biocompatibility, and high-quality factor for sensing[45, 46]. Microdroplet lasers made from various types of materials have been demonstrated for sensing in solutions and body fluids; however, not been directly on human or physiological sensing applications before. To obtain an active microlaser with biochemical sensing functions, here we developed a wearable thin film laser formed by encapsulating cholesteric liquid crystal (CLC) droplets in a flexible hydrogel thin film, as shown in Figure 1(a). Each single CLC microdroplet serves as a WGM microresonator and the lasing wavelength is determined by the gain (fluorescence dye) doped within liquid crystal molecules. The three-dimensional cross-linked hydrophilic polymer serves as the adhesive layer to allow small molecules to penetrate from human tissue to the surface of droplet laser resonators. Ascribable to the high-quality factor of the whispering gallery mode (WGM) resonator, subtle changes in the liquid crystal droplets will be amplified, resulting in a wavelength shift in the laser emission spectra, which can then be applied for sensing and monitoring metabolites.

To achieve the desired sensing functionality for lactate, glucose, and urea, CLC microdroplets doped with different photoluminescent dyes were modified with specific molecules. The hydrogels were blended with different corresponding enzymes. The working principle of the CLC microlaser is illustrated in Fig. 1b. During perspiration, the metabolites will react with the enzyme groups in the hydrogel and release certain chemicals, which will further react with the CLC microdroplets, leading to protonation and deprotonation of the modification molecules and the alignment of liquid crystal molecules. With the change of orientation of liquid crystal molecules, the lasing signal will also change owing to the fluctuation of the refractive index, indicating the concentration of the metabolites in sweat. As shown in Fig. 1c, the modified CLC microdroplets with different dyes and similar sizes are evenly distributed in the hydrogel.

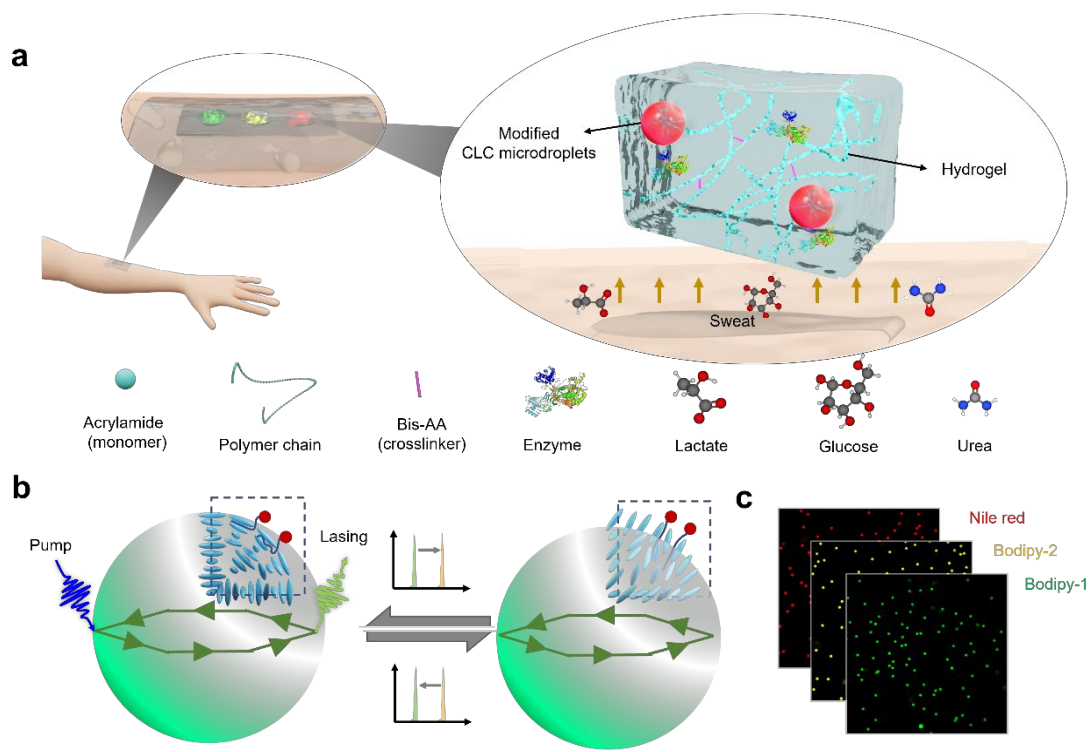

**Figure 1. Schematic concept and principle of multifunctional flexible laser based on modified CLC droplets.** (**a**) Schematic illustration of the multifunctional flexible laser for sweat sensing. (**b**) Illustration of the sensing mechanism of the modified CLC microdroplets for generating whispering gallery modes. (**c**) Captured images of luminescent CLC droplets embedded in hydrogel (different dyes were used to dop CLC droplets: Bodipy-1, Bodipy-2, and Nile Red).

## 2. Results

### 2.1 Optical characterization of modified CLC microdroplets

Modified CLC droplets were first prepared in a surfactant such as polyvinyl alcohol (PVA) solution, yielding a strong lasing emission as illustrated Fig. 2(a). Subsequently, we investigated the lasing performance of modified CLC microdroplets when transferred into a polyacrylamide (PAAm) hydrogel thin film material. Figure 2(b) shows that lasing emission could still be obtained even embedded in PAAm hydrogel. Details of the fabrication and optical setup are described in Supplementary Information Fig. S1 and S2. With increasing pump energy density, the photoluminescence intensity at ∼540 nm was amplified dramatically, manifesting the lasing action from the Bodipy-1 dye molecules. However, the lasing threshold of CLC microdroplets in PAAm hydrogel is higher (Fig. 2c), owing to the relatively higher refractive index of PAAm hydrogel as compared to PVA solution. Figure 2c shows the lasing threshold of CLC microdroplets in PAAm with different dyes (Bodipy-1, Bodipy-2, and Nile red). Bright WGM lasing emissions can be observed at the outer boundary of the microdroplets (Fig. 2d, inset), demonstrating a total internal reflection of the emitted light along the edge of the microcavity. To further explore the stability of the CLC droplet resonators in PAAm hydrogel for flexible and wearable applications, bending and temperature stability tests were carried out (Fig. S3). As shown in Fig. 2e, the size and the shape of the CLC microdroplets (Fig. 2e, inset) remained unchanged when the hydrogel film was bent from 0° to 120°, leading to the stable lasing performance. This is due to the little pressure applied on the CLC microdroplets when droplets are located at the center of the hydrogel film. The lasing spectra of CLC microdroplets in PAAm also remain stable when the temperature of the surrounding environment increases from 27 °C to 40 °C (Fig. 2f). The presence of the water in the hydrogel retains the temperature at a certain value despite the temperature change at the outer environment. Note that the observation time is only less than 10 minutes.

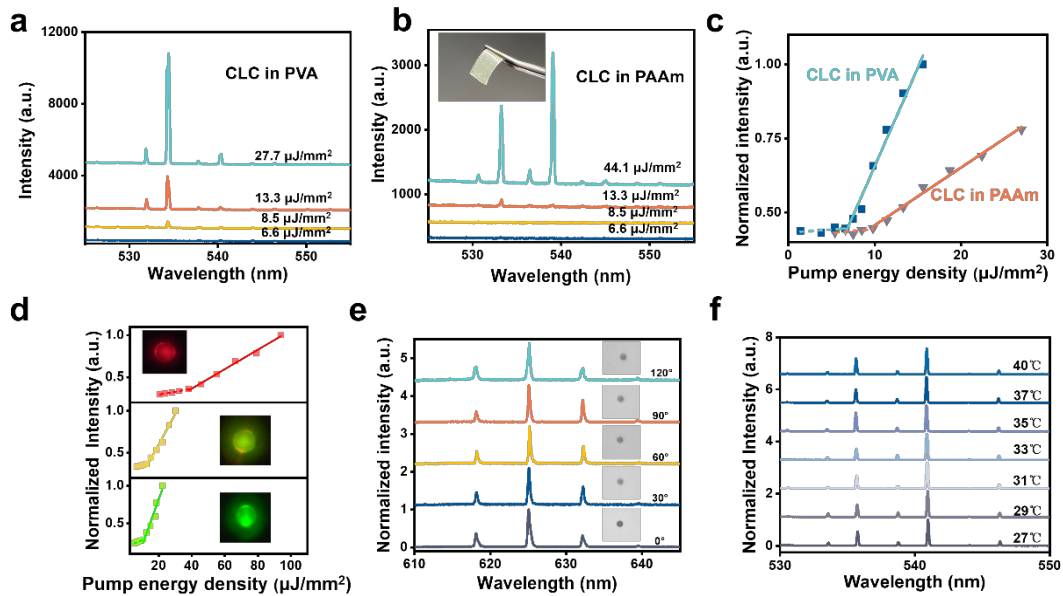

**Figure 2. The lasing performance of dye-doped CLC microdroplets in hydrogel.** (**a**) The lasing spectra of modified CLC microdroplets in PVA solution under different pump energy densities. (**b**) The lasing spectra of modified CLC microdroplets in PAAm under different pump energy densities. The inset shows an image of the device. (**c**) The lasing threshold of dye-doped CLC microdroplets in PVA and PAAm. (**d**) The lasing threshold of CLC microdroplets in PAAm when liquid crystals are doped with Bodipy-1, Bodipy-2, and Nile red, respectively. (**e**) The lasing spectra of Nile Red doped CLC microdroplets in PAAm with different bending angles (0° to 120°). Inset, the morphology of the CLC microdroplets at different bending angles. (**f**) The lasing spectra of Bodipy-1 doped CLC microdroplets in PAAm under different environmental temperatures, from 27 °C to 40 °C.

## 2.2 Lactate sensing with modified CLC microdroplets in PAAm film

Next, we explored the sensing capability of CLC droplet lasers encapsulated in PAAm hydrogel thin film. Taking advantage of the orientation shift of liquid crystal molecules during the protonation and deprotonation of changed molecules, liquid crystal droplets have been employed for sensing in many applications[46, 47]. As such, the surface of CLC droplets can be easily modified with various molecules, allowing versatile and designable functionality. To obtain lactate sensing functionality, 1-dodecanethiol was employed to align with the liquid crystal molecules and the hydrophilic thiol (Fig. S4). The lactate oxide (LOx), which is fixed into the PAAm hydrogel during the manufacturing of the flexible laser film, will assist in oxidizing lactate molecules when the lactate molecules enter the PAAm hydrogel, as illustrated in Fig. 3a. The oxidation of lactate will liberate pyruvate and $H_2O_2$ into the PAAm film, which will then oxidize the thiol (-SH) on the surface of the CLC microdroplets into sulfonyl hydroxide (-$SO_3H$). Since the polarity of -$SO_3H$ is stronger than that of -SH, the process can be seen as a kind of protonation, leading to the rotation of the liquid crystal molecules. Owing to the decrease in the refractive index which TM modes sense, the lasing wavelength of microdroplets will experience a blue shift (Fig. S5).

With the presence of lactate, obvious lasing wavelength shifts were observed as the CLC droplet lasers interact with lactate molecules (Figure 3(b)). Different lactate concentrations lead to different extents of lasing wavelength shifts. Without the presence of lactate (control group), the lasing peak remains stable with less than 0.05 nm wavelength shift. The summary of overall wavelength shifts in Fig. 3(c) revealed a linear dependence with time under various lactate concentrations (1 to 20 mM), indicating continuous enzyme catalysis and the release of $H_2O_2$. To have a fair comparison for screening analysis, we compared the sensing results under a fixed time window (6 minutes). As illustrated in Fig. 3d, wavelength shift of 0.31 nm, 0.67nm, 1.89nm, and 3.19nm was acquired after applying 1mM, 5mM, 10mM, and 20 mM lactate for 6 minutes, respectively. This range was selected according to human physiological conditions.

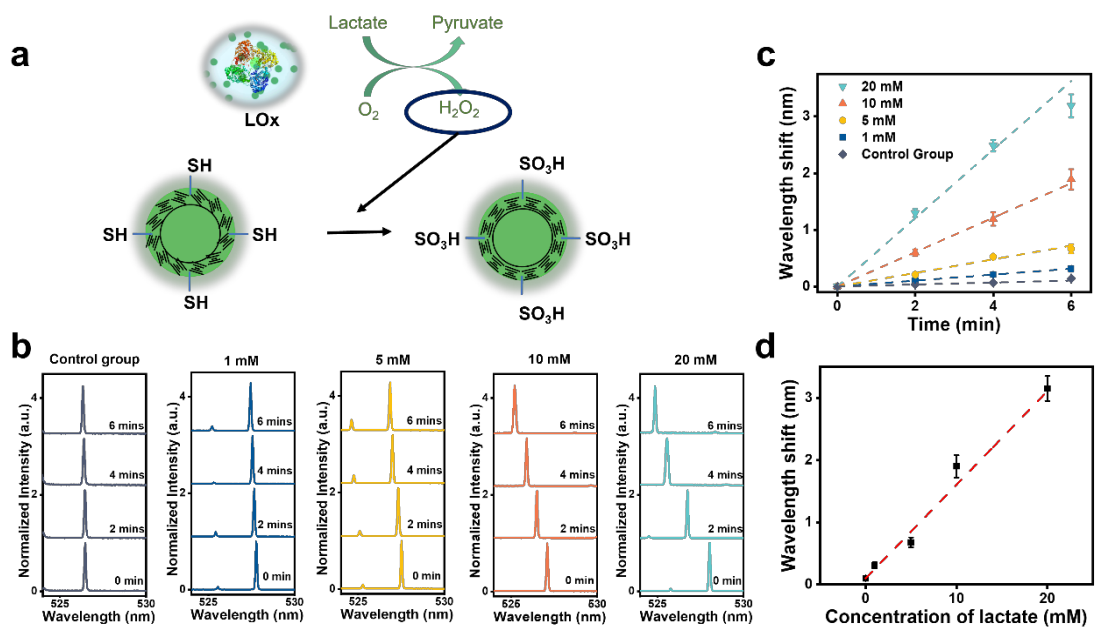

**Figure 3. Sensing principle and performance of the modified CLC microdroplet for lactate sensing.** (**a**) Working principle of the modified CLC microdroplets for lactate sensing. (**b**) Lasing spectra of CLC microdroplets in hydrogel under different lactate concentrations. The concentration of lactate used in this experiment is 1 mM, 5 mM, 10 mM, and 20 mM. (**c**) The wavelength shifts with increased lactate concentration after 2, 4, and 6 minutes. A slight blue shift (~0.05 nm) was found in the control group experiment (0 mM lactate), resulting from the bleaching of the BODIPY-1. (**d**) The wavelength shifts with increased lactate concentration under a fixed observation time (6 minutes). Error bars are obtained based on quintuplicate measurements.

## 2.3 Glucose and urea sensing with modified CLC microdroplets in PAAm film and Selectivity examination

Besides detecting lactate, we next explored the possibility of using CLC droplet lasers to detect glucose and urea in hydrogel environment. For glucose detection, a similar principle was employed owing to the generation of $H_2O_2$ during the reaction of glucose oxidization when glucose encounters glucose oxidase (GOx) (Fig. 4a). However, the modification of CLC microdroplet lasers was adjusted to fulfill the requirements for lower glucose concentration in sweat, as shown in Fig. S4. For urea detection, the PAAm hydrogel was blended with urease, while the carboxylic acid group (-COOH) was modified in order to respond to $NH_3$ (oxidation product of urea). Note that the reaction between $NH_3$ and -COOH can be seen as a process of deprotonation, which will lead to a reverse rotation of liquid crystal molecules. As a result, a red shift was discovered during the urea sensing experiment in this study.

We then applied glucose and urea to CLC microlasers and obtained the following results in Fig. 4b. The applied concentrations were selected according to the physiological conditions in human sweat (glucose: 10 μM to 200 μM; urea: 5 mM to 40 mM). Figure 4b illustrates a strong linear relationship between wavelength shift and different reaction time by applying different glucose and urea concentrations. Under a fixed observation time, a wavelength shift of 0.09 nm, 0.15nm, 0.26nm, and 0.47nm was recorded when adding 10 μM, 50 μM, 100 μM, and 200 μM glucose solution; a wavelength shift of 0.54 nm, 1.5 nm, 2.87nm, and 6.03 nm was recorded when adding 5 mM, 10 mM, 20 mM, and 40 mM urea solution.

Next, we analyzed the selectivity of CLC microdroplets in PAAm thin film, as illustrated in Fig. 4c. To conduct the selectivity experiment, we prepared lactate, glucose, and urea solutions with a concentration similar human sweat condition. Additionally, sodium chloride and ethanol were also prepared as both are significant components in sweat. As presented in Fig. 4c, significant lasing wavelength shift was only observed for the CLC microdroplets modified to detect lactate, glucose, or urea, respectively. Our results demonstrate that the modified CLC microlasers possess a good selectivity with only minimum effect on the wavelength shift. Both sodium chloride and ethanol showed a negligible impact on the wavelength shift as well.

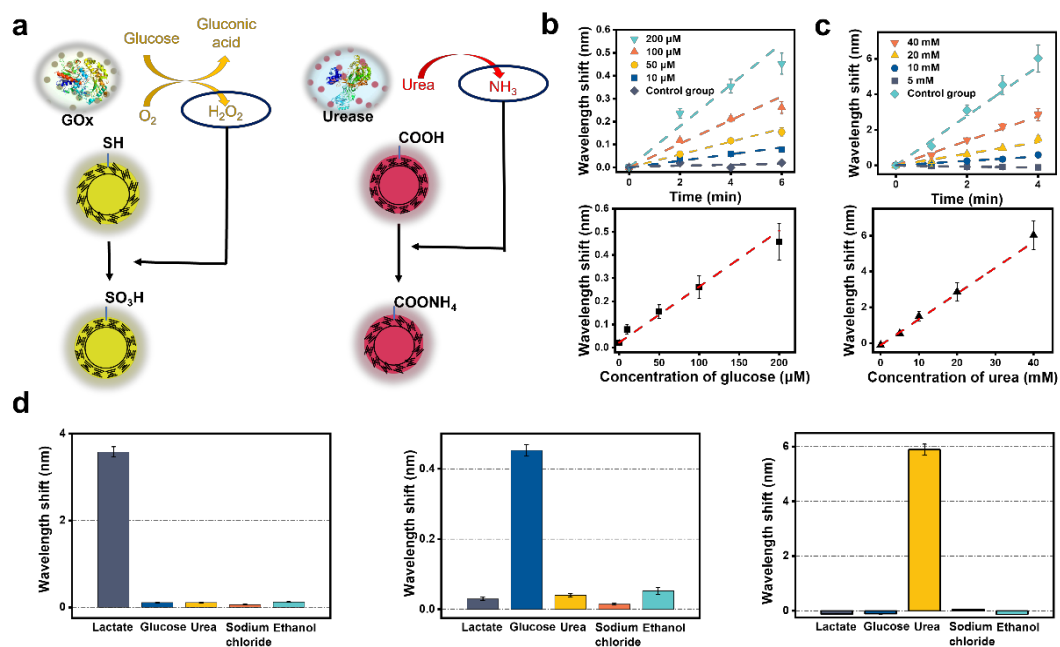

**Figure 4. Sensing principle and performance of the modified CLC microdroplet for glucose and urea sensing.** (**a**) Working principle of the modified CLC microdroplets for glucose and urea sensing. (**b**) Lasing spectra of CLC microdroplets in the hydrogel under different glucose concentration. The concentration of glucose used in this experiment is 10 μM, 50 μM, 100 μM, and 200 μM. (**c**) Lasing spectra of CLC microdroplets in the hydrogel under different urea concentration The concentration of urea used in this experiment is 5 mM, 10 mM, 20 mM, and 40 mM. (**c**) The wavelength shifts with increased lactate concentration at 2, 4, and 6 minutes. And small blue shifts in the control group were also founded in both experiments, due to the bleaching of dye (Bodipy-2 and Nile red). (**d**) The selectivity of the multifunctional flexible laser. Error bars are obtained based on quintuplicate measurements.

## 2.4 Demonstration of the modified CLC microdroplets in PAAm hydrogel film for sweat sensing

Finally, as a proof-of-concept, we demonstrate the proposed CLC microlaser can detect human sweat with multifunctionality in Figure 5. The integrated device consists of three primary pieces, as shown in Fig. 5a: a piece of medical tape, a PDMS substrate, and the lasing sensor sections, which are made up of three different types of CLC microdroplets in a PAAm hydrogel film. To detect three different metabolites, CLC microdroplets with three different lasing emission wavelengths were fabricated (doping liquid crystal with different gain/dyes: Bodipy-1, Bodipy-2, and Nile Red). The PDMS substrate served as a mold for creating the sensor components. To secure the entire apparatus to the user's skin, the medical tape was utilized. This wearable design holds great potential for future use with mobile phones for wireless connectivity and the immediate transmission of health information. The specific fabrication process is shown in Fig. S2.

Three sensors were tested independently with different methods to create proper conditions. In the context of urea sensing testing, this research endeavor aimed to investigate the wavelength shift of the modified CLC microdroplets in polyacrylamide (PAAm) hydrogel containing urease, both before and after protein intake for 2 hours. As depicted in Fig. 5b, the average wavelength shift within 6 minutes after protein consumption for 2 hours was found to be 4.16 nm, which is significantly higher than the pre-consumption value of 3.28 nm, suggesting a higher concentration of urea in sweat. This observation is consistent with the rise in urea levels in sweat after protein consumption. Similarly, a comparable approach was employed for glucose sensing, by comparing the wavelength shift before and after consuming sugar candies for 2 hours. As demonstrated in Fig. 5c, a change in the average wavelength shift was observed, with an increase from 0.28 nm to 0.41 nm, indicating a higher concentration of glucose in sweat. This phenomenon is in accordance with the rise in glucose levels in sweat after consuming sweets. With respect to lactate sensing in sweat, testing was conducted after jogging and high-intensity interval training (HIIT) for 10 minutes since lactate secretion is closely related to exercise intensity, where higher intensity results in more lactate secretion. Figure 5d illustrates the results of the two experiments, in which the wavelength shifts of the modified CLC microdroplets resonators after a 10-minute HIIT were much higher than those after a 10-minute jog (1.76 nm to 0.87 nm). Overall, the above three experiments demonstrate that our device is capable of human sweat testing.

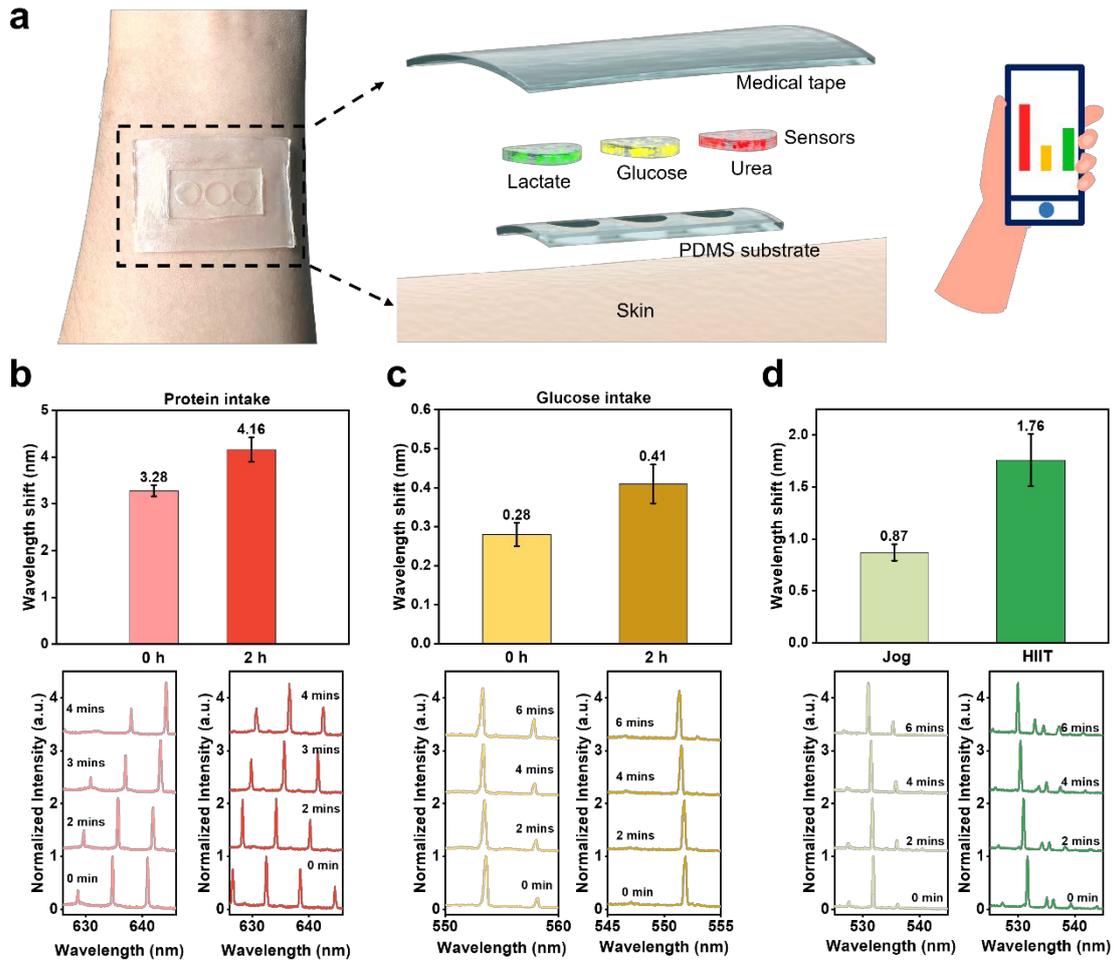

**Figure 5. Demonstration of multifunctional flexible laser integrated on human skin for direct sweat sensing.** (**a**) Structure of the multifunctional flexible laser. (**b**) Lasing wavelength shifts of CLC microdroplets in PAAm film measured before and after protein intake for 2 hours. (**c**) Lasing wavelength shifts of CLC droplets in PAAm film before and after glucose intake for 2 hours. (**d**) Lasing wavelength shifts of CLC droplets in PAAm film after jogging for 10 minutes and after HIIT for 5 minutes. Error bars are obtained based on quintuplicate measurements.

### 3. Discussion

This study explores the potential for flexible and multifunctional microlasers that can be tailored to detect various bio-signals in human sweat. By embedding modified CLC microdroplets within a PAAm hydrogel film, we were able to achieve both flexibility and physiological sensing capabilities on human skin, including lactate, glucose, and urea. According to previous clinical studies and reports, the normal range for human lactate, glucose, and urea is around 5-20 mM, 10-200 $\mu$M, and 2-10 mM respectively[22]. Hence our device fulfills the required dynamic range and it is envisioned to be applied to daily health monitoring, as it is low-cost and disposable. Furthermore, this approach can be used to detect any desired target metabolites by simply modifying the CLC microdroplets.

Here, we would like to suggest some future directions for the development of multifunctional and flexible microlasers. Firstly, the uniformity and distribution of the CLC microdroplet sensors can be improved by implementing advanced microfluidic systems or imprinted technology to form an array-like sensor with a larger sensing area. Secondly, by modifying the CLC droplets, the range of detectable biomolecules can be expanded, such as drugs secreted in sweat, pathological chemicals, and others. Thirdly, the abilities of this structure of microdroplets in the hydrogel film can be enhanced by altering the components of the droplets or the hydrogel substrate itself. Finally, although challenging, by integrating our microlaser with miniaturized laser diodes and even on-chip spectrometer[48-50], a flexible and wearable photonic chip can be created for human health monitoring.

## 4. Methods
### 4.1 Optical system setup

A microscope system (Nikon Ni2) with 20×0.3 NA objective was used to pump the microdroplets resonator and collect light. The optical pumping was performed using a pulsed nanosecond laser (EKSPLA NT230) integrated with an optical parametric oscillator, with a repetition rate of 50 Hz and a pulse duration of 5 ns. For Bodipy-1 and Bodipy-2, the excitation wavelength was set to 488 nm, while for Nile red, it was set to 532 nm. The beam diameter at the objective focal plane was approximately 20 μm. The emission laser from the microspheres was split by a beam splitter and directed to a spectrometer (Andor Kymera 328i) and a CMOS camera (Andor Zyla 5.5) for spectrum and image acquisition, respectively. A color CCD camera (DS-Fi3, Nikon) was mounted on the microscope to measure the color fluorescence images of microdroplets.

### 4.2 Fabrication of modified CLC microdroplets

Cholesteric liquid crystal (CLC) mixture was obtained by mixing 4′-Pentyl-4-biphenylcarbonitrile (Sigma, 328510) with the chiral dopant (S)-4-Cyano-4'-(2-methyl butyl) biphenyl (Tokyo Chemical Industry, C2913) in a concentration of 15 wt%. Then the modified CLC microdroplets are fabricated by sonification in polyvinyl alcohol (PVA) solution and selected by a syringe filter, whose diameters are around 12~13 μm

For lactate sensing, 10 mM Bodipy-1 (Sigma, 793728) and 0.1wt% 1-Dodecanethiol (Sigma, 471364) were doped into the CLC mixture. Bodipy-1 was used as the gain medium for lasing and the 1-Dodecanethiol served as the thiol modification. 10 μL doped CLC mixture was added to 1 mL polyvinyl alcohol (PVA) solution (1 wt%). The final mixture was treated with 1 min sonication to generate LC droplets. The obtained modified CLC microdroplets solution was filtered by using a syringe filter with a pore size of 15 μm.

For glucose sensing, a 10 mM Bodipy-2 (Sigma, 795526) and 2.5 wt% lauroyl chloride (Tokyo Chemical Industry, D0972) were doped into the CLC mixture. Bodipy-2 served as the gain medium for lasing while lauroyl chloride acted as the tailoring agent for thiol modification. Subsequently, 10 μL of the freshly prepared doped CLC mixture was added to 1 mL of polyvinyl alcohol (PVA) solution (1 wt%). The final mixture was sonicated for 1 minute to generate LC droplets, and the resulting LC droplets solution was filtered using a syringe filter with a 15 μm pore size. A solution of L-cysteine with a concentration of 1 mM was freshly prepared by dissolving L-cysteine powder (Sigma, 168149) into phosphate-buffered saline solution (pH=8.0, containing 1 wt% PVA). After centrifugation, the LC droplets were dispersed into the L-cysteine solution and incubated for 30 minutes. Finally, the modified CLC microdroplets were centrifuged and then dispersed into 1 wt% PVA for use.

For urea sensing, 15 mM Nile Red (Sigma, 72485) and 0.25 wt% of 4′-n-hexyl biphenyl-4-carboxylic acid (Alfa Aesar, B21900.03) was doped into the liquid crystal. 10 μL doped CLC mixture was added to 1 mL polyvinyl alcohol (PVA) solution (1 wt%). The final mixture was treated with 1 min sonication to generate modified CLC microdroplets. The obtained modified CLC microdroplets solution was filtered by using a syringe filter with 15 μm pore size.

## 4.3 Fabrication of the modified CLC microdroplets in PAAm hydrogel

For PAAm hydrogel precursor solution, 100 mg acrylamide and 1 mg N, N′-Methylenebis(acrylamide) (Sigma, 146072) were dissolved in 900 μL Deionized (DI) water. Furthermore, 10 mg photoinitiator was added to the solution to trigger polymerization after UV curing (Panasonic #ANUJ3500). And then the precursor solution will be modified by adding 0.1 mg lactate oxidase, 0.1 mg glucose oxidase, or 0.1 mg urea respectively for lactate, glucose, and urea sensing. All the modified CLC microdroplets solution (in 1% PVA solution) was centrifuged and then dispersed into modified PAAm hydrogel precursor solution, with specific modified CLC microdroplets corresponding to specific modified precursor solutions. The resulting solution was poured into prepared molds, and the hydrogel was fabricated after UV curing for 30 s. To remove any unreacted monomer, crosslinker, and photoinitiator, the hydrogel was soaked in DI water for 10 minutes. Two molds were utilized in this study: a cuboid mold, which was fabricated by cutting slides to a size of 10 mm * 15 mm * 1 mm, and a circular mold, which was fabricated by attaching PDMS to a slide with a diameter of 8 mm (Fig. S2). Note that CLC microdroplets are randomly distributed in the hydrogel and we choose the microdroplets lying at the middle height of the hydrogel film for the performance test and sensing application with the same size in the further experiment to get consistent and reliable results.

**4.4 Demo testing experiment.**

In the urea and glucose sensing study, the participant was instructed to walk outdoors with plastic wrap wrapped around the arm to induce perspiration. After perspiration, the device was attached to the arm for 30 seconds to collect the sweat. The lasing testing was then performed over a period of 6 minutes. After the initial walk, participants were given a standardized protein drink containing 30 g of protein for the urea sensing test, and two energy bars with a total of 54 g of sugar for the glucose sensing test. After a 2-hour period, the same testing process was repeated.

For the lactate sensing test, the participant was instructed to jog for 10 minutes with plastic wrap wrapped around their arm. After jogging, the device was attached to the arm for 30 seconds to collect the sweat. The lasing testing was then performed over a period of 6 minutes. The participant was given a 30-minute rest period before performing a 10-minute high-intensity interval training (HIIT) exercise. Finally, the same testing process was repeated.

**Acknowledgments**

This research is supported by A*STAR under its MTC IRG Grant (Project No. **M21K2c0106**). We would like to thank the lab support from Centre of Bio-Devices and Bioinformatics and Internal Grant from NTU.

**Author Contribution**

N. N. and Y. -C. C. conceived the idea and designed the experiments; N. N., X. G. and C. G. conducted the experiments; N. N., Z. Q. and Z. W. help with the analyze of the principle; G. F. provided useful advice; N. N. and Y. -C. C wrote the paper.; Y. -C. C. supervised the whole project.

**Disclosures**

All the authors declare no conflict of interest.

# References

i

1. Chen, D.; Pei, Q. Electronic Muscles and Skins: A Review of Soft Sensors and Actuators. *Chem. Rev.* **2017**, *117* (17), 11239-11268.
2. Chortos, A.; Liu, J.; Bao, Z. Pursuing prosthetic electronic skin. *Nat. Mater.* **2016**, *15* (9), 937-950.
3. Yang, J. C.; Mun, J.; Kwon, S. Y.; Park, S.; Bao, Z.; Park, S. Electronic Skin: Recent Progress and Future Prospects for Skin-Attachable Devices for Health Monitoring, Robotics, and Prosthetics. *Adv. Mater.* **2019**, *31* (48), e1904765.
4. Fan, F. R.; Tang, W.; Wang, Z. L. Flexible Nanogenerators for Energy Harvesting and Self-Powered Electronics. *Adv. Mater.* **2016**, *28* (22), 4283-4305.
5. Liu, W.; Song, M. S.; Kong, B.; Cui, Y. Flexible and Stretchable Energy Storage: Recent Advances and Future Perspectives. *Adv. Mater.* **2017**, *29* (1).
6. Liu, Y.; He, K.; Chen, G.; Leow, W. R.; Chen, X. Nature-Inspired Structural Materials for Flexible Electronic Devices. *Chem. Rev.* **2017**, *117* (20), 12893-12941.
7. Pan, J.; Zhang, Z.; Jiang, C.; Zhang, L.; Tong, L. A multifunctional skin-like wearable optical sensor based on an optical micro-/nanofibre. *Nanoscale* **2020**, *12* (33), 17538-17544.
8. Yokota, T.; Zalar, P.; Kaltenbrunner, M.; Jinno, H.; Matsuhisa, N.; Kitanosako, H.; Tachibana, Y.; Yukita, W.; Koizumi, M.; Someya, T. Ultraflexible organic photonic skin. *Sci. Adv.* **2016**, *2* (4), e1501856.
9. Choi, J. H.; No, Y. S.; So, J. P.; Lee, J. M.; Kim, K. H.; Hwang, M. S.; Kwon, S. H.; Park, H. G. A high-resolution strain-gauge nanolaser. *Nat. Commun.* **2016**, *7*, 11569.
10. Zhang, C.; Dong, H.; Zhang, C.; Fan, Y.; Yao, J.; Zhao, Y. S. Photonic skins based on flexible organic microlaser arrays. *Sci. Adv.* **2021**, *7* (31), eabh3530.
11. Karl, M.; Glackin, J. M. E.; Schubert, M.; Kronenberg, N. M.; Turnbull, G. A.; Samuel, I. D. W.; Gather, M. C. Flexible and ultra-lightweight polymer membrane lasers. *Nat. Commun.* **2018**, *9* (1), 1525.
12. Liu, W.; Liu, Y.; Yang, Z.; Xu, C.; Li, X.; Huang, S.; Shi, J.; Du, J.; Han, A.; Yang, Y.; et al. Flexible solar cells based on foldable silicon wafers with blunted edges. *Nature* **2023**, *617* (7962), 717-723.
13. Keum, C.; Murawski, C.; Archer, E.; Kwon, S.; Mischok, A.; Gather, M. C. A substrateless, flexible, and water-resistant organic light-emitting diode. *Nat. Commun.* **2020**, *11* (1).
14. Zhang, Y.-Y.; Zheng, Y.-X.; Lai, J.-Y.; Seo, J.-H.; Lee, K. H.; Tan, C. S.; An, S.; Shin, S.-H.; Son, B.; Kim, M. High Performance Flexible Visible-Blind Ultraviolet Photodetectors with Two-Dimensional Electron Gas Based on Unconventional Release Strategy. *ACS Nano* **2021**, *15* (5), 8386-8396.
15. Hsu, Y. T.; Tai, C. T.; Wu, H. M.; Hou, C. F.; Liao, Y. M.; Liao, W. C.; Haider, G.; Hsiao, Y. C.; Lee, C. W.; Chang, S. W.; et al. Self-Healing Nanophotonics: Robust and Soft Random Lasers. *ACS Nano* **2019**, *13* (8), 8977-8985.
16. Hou, Y.; Zhou, Z.; Zhang, C.; Tang, J.; Fan, Y.; Xu, F.-F.; Zhao, Y. S. Full-color flexible laser displays based on random laser arrays. *Sci. China Mater.* **2021**, *64* (11), 2805-2812.
17. Ge, K.; Guo, D.; Ma, X.; Xu, Z.; Hayat, A.; Li, S.; Zhai, T. Large-Area Biocompatible Random Laser for Wearable Applications. *Nanomaterials (Basel)* **2021**, *11* (7).
18. Zhao, C.; Liu, D.; Cai, Z.; Du, B.; Zou, M.; Tang, S.; Li, B.; Xiong, C.; Ji, P.; Zhang, L.; et al.

A Wearable Breath Sensor Based on Fiber-Tip Microcantilever. *Biosensors* **2022**, *12* (3).
19. Zhao, J.; Zhang, S.; Sun, Y.; Zhou, N.; Yu, H.; Zhang, H.; Jia, D. Wearable Optical Sensing in the Medical Internet of Things (MIoT) for Pervasive Medicine: Opportunities and Challenges. *ACS Photon.* **2022**, *9* (8), 2579-2599.
20. Zhai, T.; Chen, J.; Chen, L.; Wang, J.; Wang, L.; Liu, D.; Li, S.; Liu, H.; Zhang, X. A plasmonic random laser tunable through stretching silver nanowires embedded in a flexible substrate. *Nanoscale* **2015**, *7* (6), 2235-2240.
21. Tong, J.; Shi, X.; Wang, Y.; Han, L.; Zhai, T. Flexible plasmonic random laser for wearable humidity sensing. *Sci. China Inf. Sci.* **2021**, *64* (12).
22. Bariya, M.; Nyein, H. Y. Y.; Javey, A. Wearable sweat sensors. *Nat. Electron.* **2018**, *1* (3), 160-171.
23. Kim, J.; Campbell, A. S.; de Avila, B. E.; Wang, J. Wearable biosensors for healthcare monitoring. *Nat. Biotechnol.* **2019**, *37* (4), 389-406.
24. Koh, A.; Kang, D.; Xue, Y.; Lee, S.; Pielak, R. M.; Kim, J.; Hwang, T.; Min, S.; Banks, A.; Bastien, P.; et al. A soft, wearable microfluidic device for the capture, storage, and colorimetric sensing of sweat. *Sci. Transl. Med.* **2016**, *8* (366), 366ra165-366ra165.
25. Nyein, H. Y. Y.; Tai, L. C.; Ngo, Q. P.; Chao, M.; Zhang, G. B.; Gao, W.; Bariya, M.; Bullock, J.; Kim, H.; Fahad, H. M.; et al. A Wearable Microfluidic Sensing Patch for Dynamic Sweat Secretion Analysis. *ACS Sens.* **2018**, *3* (5), 944-952.
26. Yu, Y.; Nassar, J.; Xu, C.; Min, J.; Yang, Y.; Dai, A.; Doshi, R.; Huang, A.; Song, Y.; Gehlhar, R.; et al. Biofuel-powered soft electronic skin with multiplexed and wireless sensing for human-machine interfaces. *Sci. Robot.* **2020**, *5* (41), eaaz7946.
27. Mishra, A.; Greaves, R.; Massie, J. The relevance of sweat testing for the diagnosis of cystic fibrosis in the genomic era. *Clin. Biochem. Rev.* **2005**, *26* (4), 135-153.
28. Zafar, H.; Channa, A.; Jeoti, V.; Stojanovic, G. M. Comprehensive Review on Wearable Sweat-Glucose Sensors for Continuous Glucose Monitoring. *Sensors (Basel)* **2022**, *22* (2).
29. Qiao, L.; Benzigar, M. R.; Subramony, J. A.; Lovell, N. H.; Liu, G. Advances in Sweat Wearables: Sample Extraction, Real-Time Biosensing, and Flexible Platforms. *ACS Appl. Mater. Interfaces* **2020**, *12* (30), 34337-34361.
30. Xu, X.-Y.; Yan, B. A fluorescent wearable platform for sweat Cl− analysis and logic smart-device fabrication based on color adjustable lanthanide MOFs. *J. of Mater. Chem. C* **2018**, *6* (7), 1863-1869.
31. Bajgrowicz-Cieslak, M.; Alqurashi, Y.; Elshereif, M. I.; Yetisen, A. K.; Hassan, M. U.; Butt, H. Optical glucose sensors based on hexagonally-packed 2.5-dimensional photonic concavities imprinted in phenylboronic acid functionalized hydrogel films. *RSC Advances* **2017**, *7* (85), 53916-53924.
32. Koh, E. H.; Lee, W. C.; Choi, Y. J.; Moon, J. I.; Jang, J.; Park, S. G.; Choo, J.; Kim, D. H.; Jung, H. S. A Wearable Surface-Enhanced Raman Scattering Sensor for Label-Free Molecular Detection. *ACS Appl. Mater. Interfaces* **2021**, *13* (2), 3024-3032.
33. He, X.; Fan, C.; Luo, Y.; Xu, T.; Zhang, X. Flexible microfluidic nanoplasmonic sensors for refreshable and portable recognition of sweat biochemical fingerprint. *npj Flexible Electronics* **2022**, *6* (1).
34. Gong, C.; Sun, F.; Yang, G.; Wang, C.; Huang, C.; Chen, Y. C. Multifunctional Laser Imaging of Cancer Cell Secretion with Hybrid Liquid Crystal Resonators. *Laser & Photonics Rev.* **2022**,

*16* (8).

35. Chen, Y. C.; Tan, X.; Sun, Q.; Chen, Q.; Wang, W.; Fan, X. Laser-emission imaging of nuclear biomarkers for high-contrast cancer screening and immunodiagnosis. *Nat. Biomed. Eng.* **2017**, *1*, 724-735.

36. Fan, X.; Yun, S.-H. The potential of optofluidic biolasers. *Nat. Methods* **2014**, *11* (2), 141-147.

37. Chen, Y. C.; Chen, Q.; Fan, X. Lasing in blood. *Optica* **2016**, *3* (8), 809-815.

38. Gather, M. C.; Yun, S. H. Single-cell biological lasers. *Nat. Photon.* **2011**, *5* (7), 406-410.

39. He, L.; Özdemir, Ş. K.; Yang, L. Whispering gallery microcavity lasers. *Laser & Photonics Rev.* **2012**, *7* (1), 60-82.

40. Santiago‐Cordoba, M. A.; Cetinkaya, M.; Boriskina, S. V.; Vollmer, F.; Demirel, M. C. Ultrasensitive detection of a protein by optical trapping in a photonic‐plasmonic microcavity. *J. of Biophoton.* **2012**, *5* (8-9), 629-638.

41. He, L.; Özdemir, Ş. K.; Zhu, J.; Kim, W.; Yang, L. Detecting single viruses and nanoparticles using whispering gallery microlasers. *Nat. Nanotechnol.* **2011**, *6* (7), 428-432.

42. Wang, Y.; Zeng, S.; Humbert, G.; Ho, H. P. Microfluidic Whispering Gallery Mode Optical Sensors for Biological Applications. *Laser & Photonics Rev.* **2020**, *14* (12).

43. Yuan, Z.; Zhou, Y.; Qiao, Z.; Eng Aik, C.; Tu, W.-C.; Wu, X.; Chen, Y.-C. Stimulated Chiral Light–Matter Interactions in Biological Microlasers. *ACS Nano* **2021**, *15* (5), 8965-8975.

44. Toropov, N.; Cabello, G.; Serrano, M. P.; Gutha, R. R.; Rafti, M.; Vollmer, F. Review of biosensing with whispering-gallery mode lasers. *Light Sci. Appl.* **2021**, *10* (1), 42.

45. Wang, Z.; Zhang, Y.; Gong, X.; Yuan, Z.; Feng, S.; Xu, T.; Liu, T.; Chen, Y.-C. Bio-electrostatic sensitive droplet lasers for molecular detection. *Nanoscale Adv.* **2020**, *2* (7), 2713-2719.

46. Duan, R.; Li, Y.; Shi, B.; Li, H.; Yang, J. Real-time, quantitative and sensitive detection of urea by whispering gallery mode lasing in liquid crystal microdroplet. *Talanta* **2020**, *209*.

47. Lee, H.-G.; Munir, S.; Park, S.-Y. Cholesteric Liquid Crystal Droplets for Biosensors. *ACS Appl. Mater. & Interfaces* **2016**, *8* (39), 26407-26417.

48. Datta, I.; Chae, S. H.; Bhatt, G. R.; Tadayon, M. A.; Li, B.; Yu, Y.; Park, C.; Park, J.; Cao, L.; Basov, D. N.; et al. Low-loss composite photonic platform based on 2D semiconductor monolayers. *Nat. Photon.* **2020**, *14* (4), 256-262.

49. Cai, G.; Li, Y.; Zhang, Y.; Jiang, X.; Chen, Y.; Qu, G.; Zhang, X.; Xiao, S.; Han, J.; Yu, S.; et al. Compact angle-resolved metasurface spectrometer. *Nat. Mater.* **2023**.

50. Yang, Z.; Albrow-Owen, T.; Cui, H.; Alexander-Webber, J.; Gu, F.; Wang, X.; Wu, T.-C.; Zhuge, M.; Williams, C.; Wang, P.; et al. Single-nanowire spectrometers. *Science* **2019**, *365* (6457), 1017-1020.